\documentclass[a4paper,twocolumn]{jpconf}
\usepackage{graphicx}
\begin{document}
\title{Inelastic fingerprints of hydrogen contamination in atomic gold wire systems}

\author{Thomas Frederiksen, Magnus Paulsson, and Mads Brandbyge}

\address{MIC -- Department of Micro and Nanotechnology,
  NanoDTU, Technical University of Denmark, {\O}rsteds
  Plads, Bldg.~345E, DK-2800 Lyngby, Denmark}

\ead{thf@mic.dtu.dk}

\begin{abstract}
We present series of first-principles calculations for both pure and
hydrogen contaminated gold wire systems in order to investigate how
such impurities can be detected. We show how a single H atom or a
single H$_2$ molecule in an atomic gold wire will affect forces and
Au-Au atom distances under elongation. We further determine the
corresponding evolution of the low-bias conductance as well as the
inelastic contributions from vibrations. Our results indicate that
the conductance of gold wires is only slightly reduced from the
conductance quantum G$_0=2e^2/h$ by the presence of a single
hydrogen impurity, hence making it difficult to use the conductance
itself to distinguish between various configurations. On the other
hand, our calculations of the inelastic signals predict significant
differences between pure and hydrogen contaminated wires, and,
importantly, between atomic and molecular forms of the impurity. A
detailed characterization of gold wires with a hydrogen impurity
should therefore be possible from the strain dependence of the
inelastic signals in the conductance.
\end{abstract}

\section{Introduction}
In the late 1990s it was discovered that gold can form free-standing
single-atomic wires \cite{AgYeva.03.Quantumpropertiesof}. It was
first observed in molecular dynamics simulations of the formation of
an atomic point contact
\cite{FiLyMc.97.Atomisticsimulationof,SoBrJa.98.Mechanicaldeformationof},
and soon after also demonstrated experimentally
\cite{OhKoTa.98.Quantizedconductancethrough,YaBova.98.Formationandmanipulation}.
One of two popular techniques is typically used for creating such
atomic gold wires. By utilizing the mechanical control of a scanning
tunneling microscope (STM) to first contact a gold surface with a
gold tip and next slowly withdraw the tip such that the gold bridge
thins out, it may lead to the formation of a chain of single atoms
\cite{OhKoTa.98.Quantizedconductancethrough}. The other method is
based on the mechanically controllable break-junction (MCBJ)
consisting of a macroscopic gold wire mounted on a flexible
substrate, which is bent until the wire breaks and exposes clean
fracture surfaces \cite{YaBova.98.Formationandmanipulation}. By
controlling the bending it is possible to repeatedly form contacts
and (in some cases) to pull chains several atoms long.

These ultimate thin metallic wires are interesting for several
reasons. They are nearly ideal realizations of the perfectly
transmitting one-dimensional conductor, and have a conductance close
to the quantum G$_0=2e^2/h$ due to a single transmission channel.
Also their mechanical and chemical properties are very different
from that of bulk gold due the low coordination of chain atoms.
Further, these wires allow for studying various fundamental quantum
phenomena that are excellent for bench-marking new theoretical
models and schemes.

While gold is usually perceived as an inert material it is known
that low coordinated atoms---e.g., around surface step edges---are
more chemically active
\cite{HANO.95.WHYGOLDIS,Bahn2001,BaLoNo.02.Adsorption-inducedrestructuringof}.
Consequently it is expected that atoms arranged in a wire geometry
(with only two nearest neighbors) may be strongly reactive and hence
prone to contamination. Indeed, a substantial amount of work has
addressed issues related to the incorporation of various impurities
in atomic gold wire systems
\cite{Bahn2001,BaLoNo.02.Adsorption-inducedrestructuringof,LeGaRo.02.Originofanomalously,
CsHaMi.03.Fractionalconductancein,
Nodada.03.Effectofimpurities,SkSi.03.Stabilityofgold,
BaHaSc.04.Hydrogenweldingand,Frdada.04.Effectofimpurities,
LeRoUg.04.Contaminantsinsuspended,LeRoUg.05.Commenton"Contaminants,
CsHaMi.06.Pullinggoldnanowires,Nodada.06.Oxygenclampsin,
ThMaBr.06.Oxygen-enhancedatomicchain,JePeOr.06.Hydrogendissociationover}.
One motivation for some of these studies was the anomalously large
Au-Au distances (as long as 4 {\AA}) which were directly observed by
Ohnishi et al.~\cite{OhKoTa.98.Quantizedconductancethrough} using
transmission electron microscopy (TEM). To account for this
observation researchers have therefore proposed that various
light-weight impurities could be present in the wire, because these
are difficult to detect with TEM due to their low contrast. Bahn et
al.~\cite{Bahn2001,BaLoNo.02.Adsorption-inducedrestructuringof}
investigated the interaction of the diatomic molecules CO, N$_2$,
and O$_2$ with an infinite gold wire model employing density
functional theory (DFT), and suggested that oxygen is a likely
candidate to form stable wires with Au-Au distances of more than 3.8
{\AA}. Later Novaes et
al.~\cite{Nodada.03.Effectofimpurities,Frdada.04.Effectofimpurities}
and Legoas et al.~\cite{LeGaRo.02.Originofanomalously,
LeRoUg.04.Contaminantsinsuspended,LeRoUg.05.Commenton"Contaminants}
examined several other impurity candidates with DFT and disputed
whether H or C in fact is the most realistic contaminant accounting
for the long bond length. Independently, Skorodumova and Simak also
presented DFT-based calculations of gold wires with hydrogen that
showed long Au-Au distances \cite{SkSi.03.Stabilityofgold}.

Beside these structural considerations the implications of hydrogen
on the electronic transport properties of atomic gold wires have
also been addressed both theoretically
\cite{BaHaSc.04.Hydrogenweldingand,JePeOr.06.Hydrogendissociationover}
and experimentally
\cite{CsHaMi.03.Fractionalconductancein,CsHaMi.06.Pullinggoldnanowires}.
Whereas these studies generally provide evidence that hydrogen
adsorbs on the wire and possibly dissociates, the details of the
atomic arrangement are still not yet fully understood. For instance,
conclusive evidence is missing of whether the atomic or the
molecular form of hydrogen is the preferred configuration.

In a similar way that molecular hydrogen in a platinum contact has
been characterized by means of vibrational spectroscopy
\cite{SmNoUn.02.Measurementofconductance,DjThUn.05.Stretchingdependenceof},
we here present for the first time theoretical predictions for the
inelastic conductance signals of atomic gold wires influenced by the
presence of hydrogen. We consider a setup with either a single H
atom or a single H$_2$ molecule incorporated in the middle of a
short gold wire suspended between bulk gold electrodes. For
comparison we also present the inelastic transport calculations for
a pure gold wire system, for which the inelastic signals have
previously been investigated \cite{AgUnRu.02.Onsetofenergy,
FrBrLo.04.InelasticScatteringand}. We find that by studying the
inelastic signals of the gold wire formation in a hydrogen
atmosphere it is possible---under certain conditions which we
describe---to detect specific vibrational modes related to hydrogen.
In particular, our results further suggest how to differentiate
between atomic and molecular configurations.

\section{Theory}

\begin{figure}[t]
\centering
\includegraphics[width=0.68\textwidth]{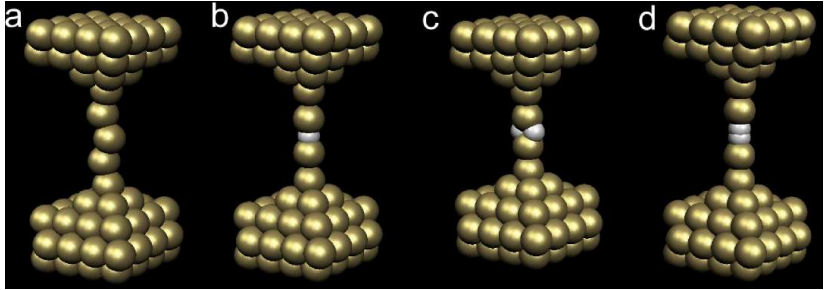}
\hfill
\begin{minipage}[b]{0.30\textwidth}\caption{\label{fig:structures} \mbox{(Color online)}
Supercells modeling (a) pure gold wires and wires contaminated with
(b) an H atom or (c-d) an H$_2$ molecule. The characteristic
electrode separation $L$ is measured between the second-topmost
surface layers. }\end{minipage}
\end{figure}

To calculate the inelastic transport properties of some atomic-scale
junction we have developed a scheme based on a combination of DFT
and non-equilibrium Green's functions (NEGF) \cite{Frederiksen2006}.
The structural properties are studied using the standard DFT
\textsc{Siesta} package \cite{SoArGa.02.SIESTAmethodab} and the
elastic conductance calculated with \textsc{TranSiesta}
\cite{BrMoOr.02.Density-functionalmethodnonequilibrium}. The
vibrations are determined by diagonalizing the dynamical matrix
extracted from finite differences and the inelastic contribution to
the conductance evaluated according to the method presented in
Ref.~\cite{PaFrBr.05.Modelinginelasticphonona}.

We consider the periodic supercell representations shown in
Fig.~\ref{fig:structures}. The electrodes are modeled by a slab
containing five Au(100) atomic layers in a $4\times 4$
representation, and the gold wire is suspended between two pyramidal
bases that connects to the electrode surfaces. The characteristic
electrode separation $L$ is measured between the second-topmost
surface layers since we relax both the wire, the pyramids, and the
first surface layers (which hence deviates on the decimals from the
bulk values). The pure gold wire setup contains 5 wire atoms, from
which we generate the contaminated structures by replacing the
middle Au atom by either a single H atom or a single H$_2$ molecule.
The corresponding calculations with \textsc{Siesta} are performed
using a single-zeta plus polarization (SZP) basis set for the Au
atoms and a split-valence double-zeta plus polarization (DZP) basis
set for the H atoms (determined using a confining energy of 0.01
Ry), the generalized gradient approximation (GGA) for the
exchange-correlation functional, a cutoff energy of 200 Ry for the
real-space grid integrations, and the $\Gamma$-point approximation
for the sampling of the three-dimensional Brillouin zone. The
interaction between the valence electrons and the ionic cores are
described by standard norm-conserving pseudo-potentials.

\section{Results}

\begin{figure}[t]
\centering
\includegraphics[width=0.67\textwidth]{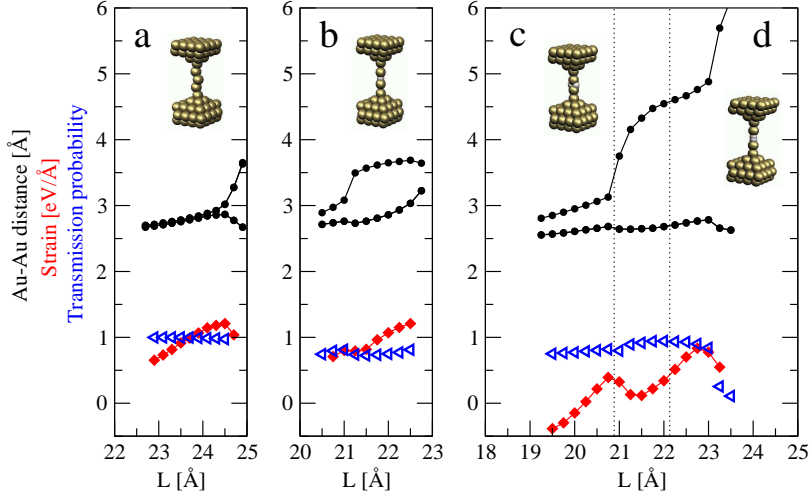}\hfill
\begin{minipage}[b]{0.3\textwidth}\caption{\label{fig:transmission} (Color
online) Mechanical and \mbox{electronic} properties of (a) pure gold
wires and wires contaminated with (b) an H atom or (c-d) an H$_2$
molecule. Black dots indicate the Au-Au distances between wire atoms
(in units of {\AA}), red squares the external force on the supercell
(in units of eV/{\AA}), and blue triangles the elastic transmission
probability at the Fermi energy.}\end{minipage}
\end{figure}

We relax the supercells under varying electrode separation $L$ to
characterize the junction as it is mechanically manipulated. The
resulting Au-Au distances between the wire atoms are shown in
Fig.~\ref{fig:transmission} with black dots. For the pure Au wire
the bond lengths gradually increase from around 2.67 {\AA} at
$L=22.70$ {\AA} (the zigzag wire depicted in
Fig.~\ref{fig:structures}a) to 2.86 {\AA} at $L=24.30$ {\AA}; beyond
this point the wire dimerizes and break. When a hydrogen impurity is
introduced the adjacent Au-Au bond becomes slightly longer than the
rest. With a single H atom in a short wire the impurity prefers to
sit to the side. As the wire is elongated to around $L=21.50$ {\AA}
the impurity moves into the center of the wire
(Fig.~\ref{fig:structures}b) resulting in an Au-Au distance larger
than 3.6 {\AA}. With H$_2$ in a short wire the impurity sits in a
transverse configuration, cf.~Fig.~\ref{fig:structures}c. At
$L=21.00$ {\AA} it begins to tilt under elongation and reaches a
straight Au-H-H-Au bridge configuration around $L=22.00$ {\AA},
cf.~Fig.~\ref{fig:structures}d. This crossover region is marked in
Fig.~\ref{fig:transmission}c-d by dotted vertical lines. Just before
breaking the Au-Au distance becomes as large as 4.9 {\AA}.

By studying how the total energy changes as the electrode separation
increases we can numerically evaluate the force on the supercell.
This is indicated in Fig.~\ref{fig:transmission} by red squares. We
generally find that it requires an external restoring force to
prevent contraction of the wires. However, for the short H$_2$
configurations this force is negative indicating the existence of a
stable situation around $L=20.00$ {\AA}. From these curves we get an
idea of the break force---defined as the maximal force under the
elongation process---which is of the order 1.2 eV/{\AA} for the pure
and single H contaminated systems, but noticeably lower in the H$_2$
case (around 0.8 eV/{\AA}).

The elastic transmission probability at the Fermi energy
$T(\varepsilon_F)$, which describes the low-temperature zero-bias
conductance via $G=\mathrm{G}_0\, T(\varepsilon_F)$, is also shown
in Fig.~\ref{fig:transmission} with blue triangles. Whereas the pure
Au wire has a conductance of (0.98-1.00)G$_0$ depending on the
length, the case of a single H atom lowers the conductance to
(0.73-0.81)G$_0$ and an H$_2$ molecule the conductance to
(0.76-0.94)G$_0$. In an experiment it may thus be difficult to
differentiate among these configurations based on a measurement of
the zero-bias conductance only.\footnote{We note that our findings
are slightly different from that of
Ref.~\cite{JePeOr.06.Hydrogendissociationover}, but differs
significantly from Ref.~\cite{BaHaSc.04.Hydrogenweldingand} that
ascribes less than 0.25 G$_0$ to a gold wire contaminated with an H
atom or an H$_2$ molecule.}

\begin{figure}[t]
\centering
\includegraphics[width=0.90\textwidth,viewport=0 28 600 350,clip]{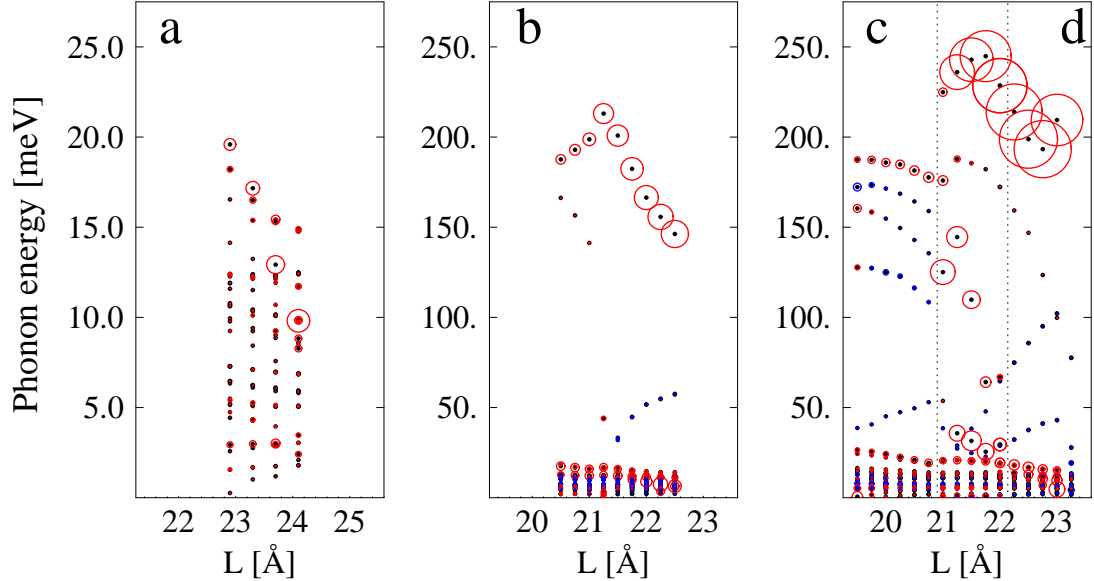}
\caption{\label{fig:inelastic-signals} \mbox{(Color online)}
Inelastic signals in the nonlinear conductance for (a) pure gold
wires and wires contaminated with (b) an H atom or (c-d) an H$_2$
molecule. The black dots mark vibrational modes at the corresponding
threshold voltages. The surrounding red (blue) circles represent
with their area the conductance decrease (increase) as observed in
an experiment.}
\end{figure}

If one instead investigates the inelastic signals we find
significant differences between the systems. Our results from a
vibrational analysis are summarized in
Fig.~\ref{fig:inelastic-signals} for which all the atoms in the
contact between the surface layers were considered to be active. The
existence of a vibrational mode is marked with a black dot at the
vibrational threshold and a corresponding decrease (increase) in the
conductance is indicated with the area of a surrounding red (blue)
circle. The pure gold wires have phonon energies in a region
comparable with the phonon density of states in bulk gold, i.e., up
to around 25 meV. A single dominant conductance decrease is seen in
Fig.~\ref{fig:inelastic-signals}a. This signal, caused by the
alternating bond length (ABL) longitudinal phonon
\cite{AgUnRu.02.Onsetofenergy,FrBrLo.04.InelasticScatteringand},
strengthens with elongation of the wire while the mode frequency
softens.

This picture is changed by the presence of light-weight impurities,
as seen from \mbox{Fig.~\ref{fig:inelastic-signals}b-d}, because
they contribute to the vibrational spectrum with new modes that lie
well above the gold phonon band. With a single H atom our
calculations predict a significant inelastic signal in the range
150-220 meV corresponding to movement of the impurity along the wire
axis. Comparatively, in the case of H$_2$ we have one inelastic
signal around 180-250 meV due to the internal H$_2$ stretch mode,
but find further two active modes in the range 25-150 meV occurring
only when the H$_2$ molecule appears in a tilted configuration
(marked by the dotted lines in Fig.~\ref{fig:inelastic-signals}c-d).
These additional modes have a transverse component and are
unambiguous indications for the H$_2$ configuration.

\section{Conclusions}
It may experimentally be difficult to determine if an atomic gold
wire contains a hydrogen impurity without measuring the inelastic
signals. We find that the low-bias conductance and the break force
of the chains are generally very similar for both pure and H or
H$_2$ contaminated wires, cf.~Fig.~\ref{fig:transmission}. However,
the inelastic conductance signals enable us to separate the
different geometries from each other. In a pure gold wire there is
generally one dominant inelastic conductance decrease which
strengthens in magnitude and decreases in threshold voltage as the
wire is elongated. This signal is caused by the ABL longitudinal
phonon. Similar signals (below 25 meV) can also be seen for the
hydrogen-contaminated wires (Fig.~\ref{fig:inelastic-signals}b-c)
reflecting that active modes involving the gold atoms survive. On
the other hand, hydrogen induces new inelastic signals at much
higher phonon energies. In the case of a single H atom (H$_2$
molecule) our calculations predict a signal approximately at 150
(200) meV just before the wire breaks. Another diversity discussed
above is the fact that two additional active modes may be detectable
if H$_2$ sits in a tilted configuration. These differences can
possibly be used to differentiate between the H and H$_2$
configurations.

\ack The authors acknowledge fruitful collaborations with
Antti-Pekka~Jauho and Nicol\'as Lorente, and thank Nicol\'as
Agra\"it for many interesting discussions related to the gold wire
experiments. Computational resources were kindly provided by the
Danish Center for Scientific Computing.

\section*{References}

\end{document}